\documentclass[slac_one]{revtex4}
\usepackage{graphicx}
\usepackage{fancyhdr}
\begin{document}
\newcommand{\hee}{\mbox{$e^+e^- \rightarrow ZH \rightarrow e^+e^-X(\gamma)$}}
\newcommand{\zee}{\mbox{$ZH \rightarrow e^+e^-X(\gamma)$}}
\newcommand{\zmm}{\mbox{$ZH \rightarrow \mu^+\mu^-X(\gamma)$}}
\newcommand{\hmm}{\mbox{$e^+e^- \rightarrow ZH \rightarrow \mu^+\mu^-X(\gamma)$}}
\newcommand{\smur}{\mbox{$e^+e^- \rightarrow \tilde{\mu}_R^+ \tilde{\mu}_R^-$}}
\newcommand{\munu}{\mbox{$\tilde{\mu}_R^{\pm} \rightarrow \mu^{\pm} \tilde{\chi}_1^0$}}
\newcommand{\slep}{\mbox{$e^+e^- \rightarrow \tilde{\ell}_R^+ \tilde{\ell}_R^-
                         \rightarrow \ell^+ \ell^- \tilde{\chi}_1^0 \tilde{\chi}_1^0$}}

\title{{\small{2005 International Linear Collider Workshop - Stanford,
U.S.A.}}\\ 
\vspace{12pt}
Impact of Tracker Design on Higgs and Slepton Measurements
} 

%

\author{Hai-Jun Yang$^*$,~~ Keith Riles$^\dagger$
\\($^*$ yhj@umich.edu, $^\dagger$ kriles@umich.edu)
}


\affiliation{Department of Physics, University of Michigan, Ann Arbor, MI 48109-1120, USA}




\begin{abstract}

We have studied the impact of charged track resolution on Higgs mass and
production cross section measurement in the process 
$e^+e^- \rightarrow Z^0 H, Z^0 \rightarrow \ell^+\ell^-,
H \rightarrow X$ for Higgs masses between 120 and 160 GeV, 
and on smuon mass measurement using smuon pair production for
three different mass pairs,
assuming the International Linear Collider(ILC) is operated
at 500 GeV center of mass energies (CME) with integrated
luminosities of 500 $fb^{-1}$. The effect of initial state radiation (ISR), beamstrahlung and
beam energy spread on Higgs and smuon mass measurement are also 
estimated.
Using fast Monte Carlo simulations of
the 2001 North American baseline detector designs (LD and SD),
we find that the ISR and beamstrahlung have significant impact on 
Higgs and smuon mass measurement; Charged track momentum resolution affect
Higgs mass significantly with better track performance yielding better
Higgs mass resolution and precision for the track momentum resolution improvement up to
a factor of about 5; Track momentum resolution has little effect on the measurement precision of
the Higgsstrahlung cross section, the branching ratio of $H \rightarrow C\bar{C}$,
the smuon and neutralino masses; Beam energy spread better than about 0.2\% has little effect on 
Higgs, smuon and neutralino masses; The SD detector provides a more 
accurate measurement than the LD of the Higgs mass.

\end{abstract}

\maketitle

\section{Introduction}

Precise measurements of Higgs boson mass and cross section are important
goals in the future high energy $e+e-$ linear collider experiments
~\cite{nlctdr,teslatdr}. Precision electroweak data in the framework of the
Standard Model predict the mass of the Higgs boson, allowing a crucial
cross check of electroweak symmetry breaking models if and when
the Higgs boson is discovered. 
In addition, measuring the production cross section accurately allows
determination of absolute Higgs decay branching ratios.
Another important goal of the ILC is to discover light supersymmetric particles, 
such as sleptons, and measure their masses and production cross sections etc.
Here we explore the accuracy with which the Higgs and Higgsstrahlung cross section
can be determined at a linear collider operated at 500 GeV
CME, considering Higgs masses between 120 and 160 GeV
with integrated luminosities of 500 $fb^{-1}$. For smuon and neutralino mass measurement,
three different mass pairs with high(47 GeV), medium(28 GeV) and low (6 GeV)
mass differences are considered using right-hand smuon pair production, with each smuon decaying
into muon and neutralino.
For this study, we use the 2001 North American baseline detector designs 
(LD for ``Large'' and SD for ``Silicon'').


\section{Higgs Property Measurement}

The Higgs mass can be simply determined assuming recoil in the process
$e^+e^- \rightarrow Z^0 H, Z^0 \rightarrow \ell^+\ell^-, H \rightarrow X$ 
($\ell=e,\mu$). 
The recoil mass is defined as:
$M_H^{recoil} = \sqrt{s - 2\sqrt{s}\cdot E_{\ell^+\ell^-} + M^2_{\ell^+\ell^-}}$,
where $s$ is the CME squared, $E_{\ell^+\ell^-}$ is the energy 
of the lepton pair from $Z^0$ decay, 
while $M_{\ell^+\ell^-}$ is the pair's invariant mass.
The main backgrounds of this analysis are $e^+e^- \rightarrow Z^0 Z^0$, $W^+W^-$, 
but other sources of contamination, including Bhabha events, dimuon and two photos
events are also investigated. 

\subsection{Event Selection}

Monte Carlo events in this analysis were generated by the Pandora(V2.2)-Pythia(V3.1)
package~\cite{pandora,pythia} with latest patches which includes initial state radiation, beamsstrahlung,
beam energy spread (full width is 0.11\%), hadron fragmentation and final state QCD/QED 
radiation. In addition, the electron beam is polarized to $-$85\%. 
The Java Analysis Studio(JAS)~\cite{jas} package was used 
to analyze fast detector simulation events, assuming the LDMAR01 and SDMAR01 baseline 
detectors~\cite{nlctdr}.

These studies are performed for a Linear Collider operated at CME
 of 500 GeV with integrated luminosities of 500$fb^{-1}$ each,
assuming Higgs mass between 120 and 160 GeV.
Events are selected using a cut-based approach, according to the following criteria 
~\cite{mhiggs,trackres}:

(1) A candidate lepton must have an energy greater than 10 GeV

(2) The polar angle of a lepton must satisfy $|cos\theta_e|<0.9$

(3) There must be at least 2 lepton candidates in the event

(4) The invariant mass of the lepton pair must lie within 5 GeV of the $Z^0$ mass

(5) The polar angle of two-lepton system must lie in the barrel region, 
    $|cos\theta_{e^+e^-}|<0.6$

(6) The opening angle between the two leptons should satisfy $|cos\theta_{e^+\leftrightarrow e^-}|>-0.7$

(7) The energy of the most energetic photon should be less than 100 GeV.

Cut (5) is used to suppress $Z^0 Z^0$ background, while Cut (6) rejects background from $W^+W^-$,
Cut (7) removes events with one energetic photon from $Z\gamma$.
The selection efficiency for signal is about 55-57\% for $\sqrt{s}=500$ GeV, as
listed in TABLE ~\ref{table:eff}.
The signal efficiency using the same selection cuts is lower, 48-50\%, at a 350 GeV machine, 
mainly because the lower Lorentz boost of the leptons from $Z^0$ decay leads
to a larger average opening angle. 
The major remaining background with the above selection cuts is $Z^0 Z^0$ events for which the selection 
efficiency is about 1\%. For the $Z^0 Z^0$ cross section of $475 \pm 3.4$ fb, 
about 2400  $Z^0 Z^0$ events passed the selection cuts assuming the integrated
luminosity of 500 $fb^{-1}$. The corresponding Higgs events in electron channel varies from 
about 570 to 650 for Higgs mass between 120 GeV and 160 GeV, with lower Higgs mass yielding 
more signal events because of larger production cross section.

\begin{table}[h]
\begin{center}
{\large
\begin{tabular}{|c|c|c|c|c|c|c|c|c|c|c|} \hline\hline
$m_{higgs}$ & Cross Section & \multicolumn{4}{|c|} {LD (GeV)} & \multicolumn{4}{|c|} {SD (GeV)} \\ \hline 
 (GeV)      &  (fb) &  Eff(\%) & $\sigma_{m_H}$ & $\Delta_{m_H}$ & Xsection(fb) &  Eff(\%) & 
$\sigma_{m_H}$ & $\Delta_{m_H}$ & Xsection(fb) \\ \hline
120 & 2.34 $\pm$ 0.015  & 55.3 & 7.2 & 0.34 & 2.34$\pm$0.167 & 55.3 & 5.4 & 0.31 & 2.34$\pm$0.164  \\  \hline
140 & 2.15 $\pm$ 0.022  & 56.4 & 6.2 & 0.34 & 2.15$\pm$0.144 & 56.4 & 4.8 & 0.28 & 2.15$\pm$0.142  \\  \hline
160 & 2.01 $\pm$ 0.032  & 56.6 & 4.6 & 0.34 & 2.01$\pm$0.131 & 56.7 & 3.7 & 0.27 & 2.01$\pm$0.129  \\  \hline
\end{tabular}
}\end{center}
\caption{\label{table:eff}
Selection efficiencies, Higgs mass resolution, precision and cross section of 
$e^+e^- \rightarrow Z H \rightarrow e^+ e^- X$ for two baseline
detectors, assuming ILC is operated at 500 GeV CME with Higgs masses
between 120 and 160 GeV, integrated luminosity of 500 $fb^{-1}$.
}
\end{table}

\subsection{Higgs Mass and Cross Section Measurement}

The Higgs mass resolutions are obtained by fitting the peak region of $Z^0$ recoil mass spectra using a
Gaussian distribution. A binned $\chi^2$ fit is performed with two free parameters, the Higgs mass and 
width. The $Z^0$ recoil mass resolution is about 4-7 GeV at $\sqrt{s}=500$ GeV for the two baseline
detectors, with better resolution obtained for the SD detector. This analytic fitting technique
yields slightly larger measured mass because of the ISR and beamstrahlung. This bias can be evaluated
and corrected by fitting both data and Monte Carlo events with the same techniques. The Higgs mass precision
obtained using this technique is about 0.6-1.2 GeV at $\sqrt{s}=500$ GeV, with lower Higgs masses yielding larger
uncertainties because of higher background contamination. 

In order to improve the mass precision, a second
fitting technique is used, based on two Monte Carlo samples generated using same Higgs mass hypothesis.
One sample is treated as ``data'', the other is Monte Carlo  template, and the recoil mass distribution
from ``data'' sample fitted to the Monte Carlo template using a binned $\chi^2$. The free parameters are
the mass and the cross section. This method automatically corrects for biases due to any effects correctly
modeled in the Monte Carlo simulation, and avoids degraded precision from non-optimal analytic modeling
of resolution. 

The precision of Higgs mass measured using the Monte Carlo template is about 0.3 GeV at 
$\sqrt{s}=500$ GeV, with the SD detector slightly better than LD, as shown in Table ~\ref{table:eff}.
The cross sections of the Higgsstrahlung signal measured for the Higgs mass between 120 and 160 GeV are shown
in Table ~\ref{table:eff} and Figure~\ref{fig:mh}.e. The relative precision of the cross section 
is determined to be 6.4-7.2\% at $\sqrt{s}=500$ GeV for both baseline detectors and nominal integrated
luminosity of 500 $fb^{-1}$. Slightly better precision is obtained by the SD design.

\subsection{Impact of Track Momentum Resolution on Higgs Mass and Cross Section}

\begin{figure}
{\scalebox{0.4}{\includegraphics{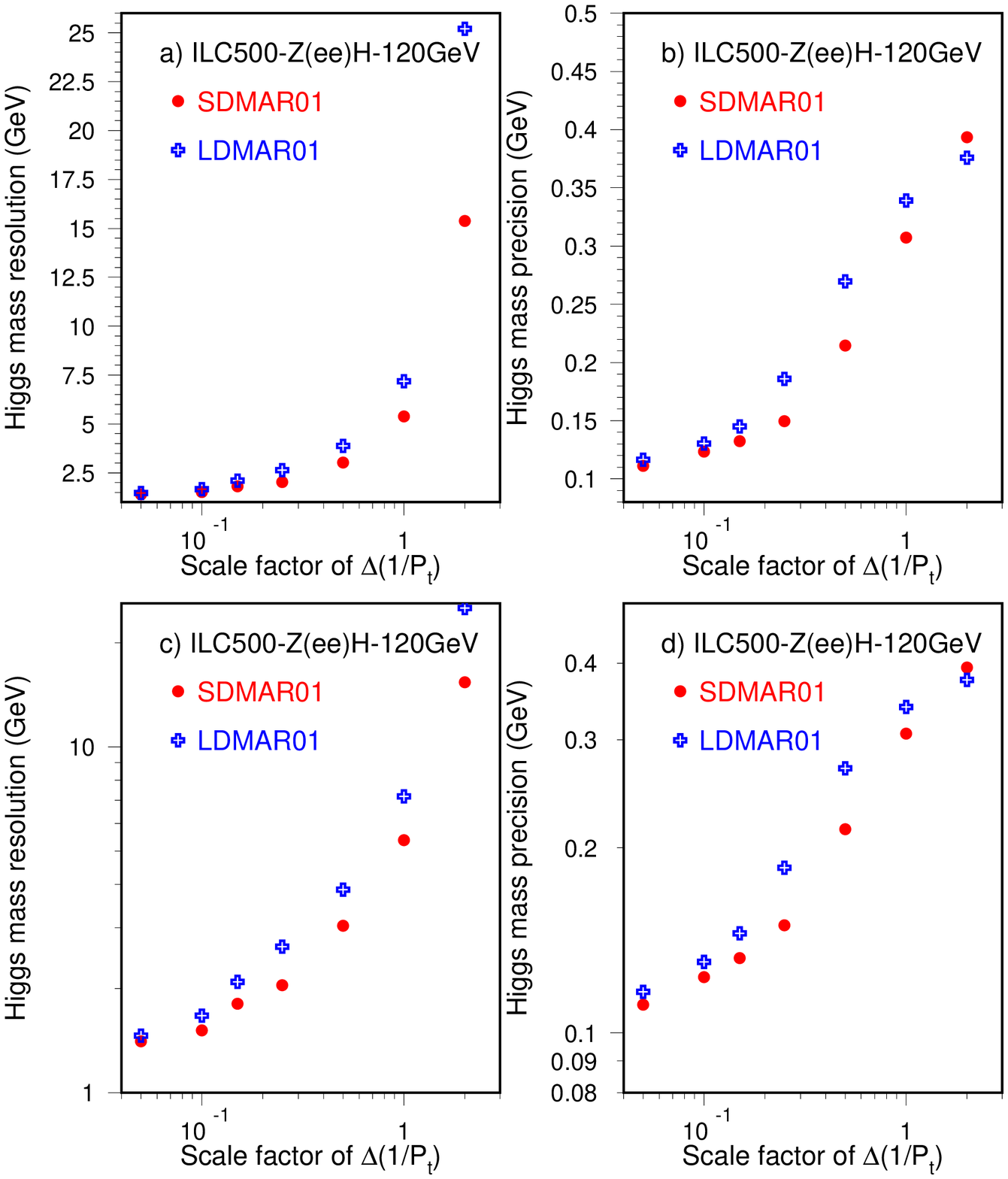}}}
{\scalebox{0.4}{\includegraphics{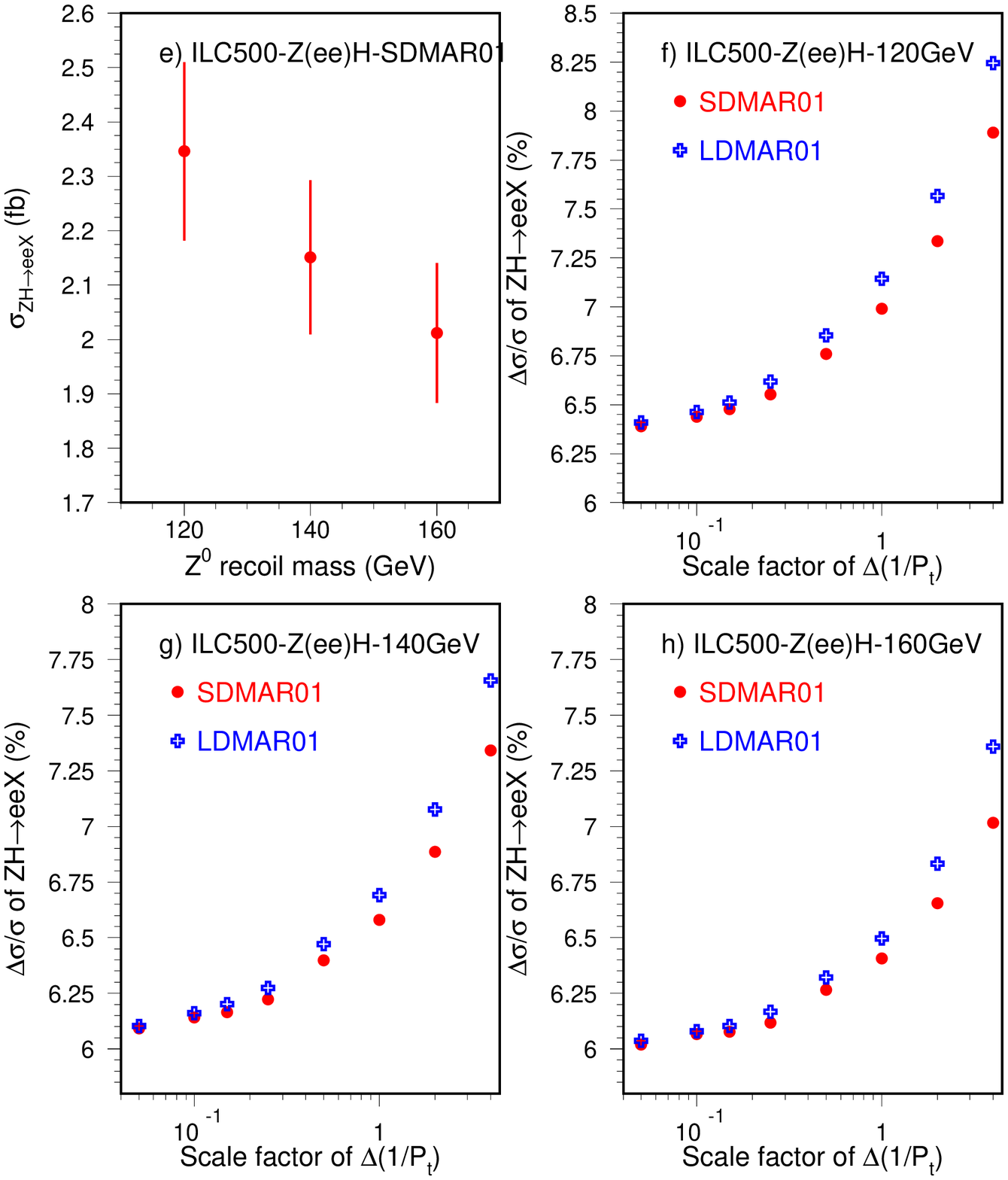}}}
\caption{\label{fig:mh} Dependence of Higgs mass resolution (a, c-logarithm scale), 
precision (b, d-logarithm scale) for 120 GeV Higgs and cross section (f-120 GeV, g-140 GeV , h-160 GeV Higgs) 
on various track momentum resolution, e for the measured cross sections of the Higgsstrahlung
versus the Higgs mass hypothesis between 120 GeV and 160 GeV with statistical errors.
}
\end{figure}

In order to quantify the effect of charged track momentum resolution on Higgs mass and cross section measurement,
the track momentum resolution is re-scaled by a factor ranging from 0.05 to 4. The results are shown in Figure 
\ref{fig:mh}.
The Higgs mass resolution and precision are improved by re-scaling the factor of track
momentum resolution down to about 0.2, the purity ($N_{signal}/(N_{signal}+N_{background})$) and 
significance ($N_{signal}/\sqrt{N_{background}}$) of the Higgsstrahlung signal are also saturated
around that point. The cross section is insensitive to charged track momentum resolution, with only 
about 10\% improvement for a the track momentum resolution improvement by a factor of 10.

\subsection{Impact of Track Momentum Resolution on Branching Ratio of $H \rightarrow c\bar{c}$}

The major background sources for $H \rightarrow C\bar{C}$ are $Z^0Z^0$ and $Z^0H^0$ with
 $Z^0 \rightarrow c\bar{c}, b\bar{b}$, or $H \rightarrow b\bar{b}, W^+W^-, gg$ etc. 
The branching ratio of $H \rightarrow b\bar{b}, W^+W^-$ strongly depends on the Higgs mass hypothesis.
The expected branching ratio of $H \rightarrow c\bar{c}$ is 2.8\% and 1.4\% for 120 GeV and 140 GeV Higgs
mass, respectively. In order to estimate the impact of track momentum resolution on the branching
ratio of $H \rightarrow c\bar{c}$, we assume 50\% c-tagging efficiency, with efficiencies of b quark
and uds quark are 4.4\% and 0.5\%, respectively. The relative precision of the $H \rightarrow c\bar{c}$
branching ratio improves only 5-10\% if we improve the track momentum resolution by a factor of 10.
For integrated luminosity of 1000 $fb^{-1}$, the relative precision is about 39\% for 120 GeV Higgs 
and 64\% for 140 GeV Higgs with the nominal track resolution.

\subsection{Effect of ISR, Beamstrahlung and Beam Energy Spread on $Z^0$ Recoil Mass}

\begin{figure}
{\scalebox{0.35}{\includegraphics{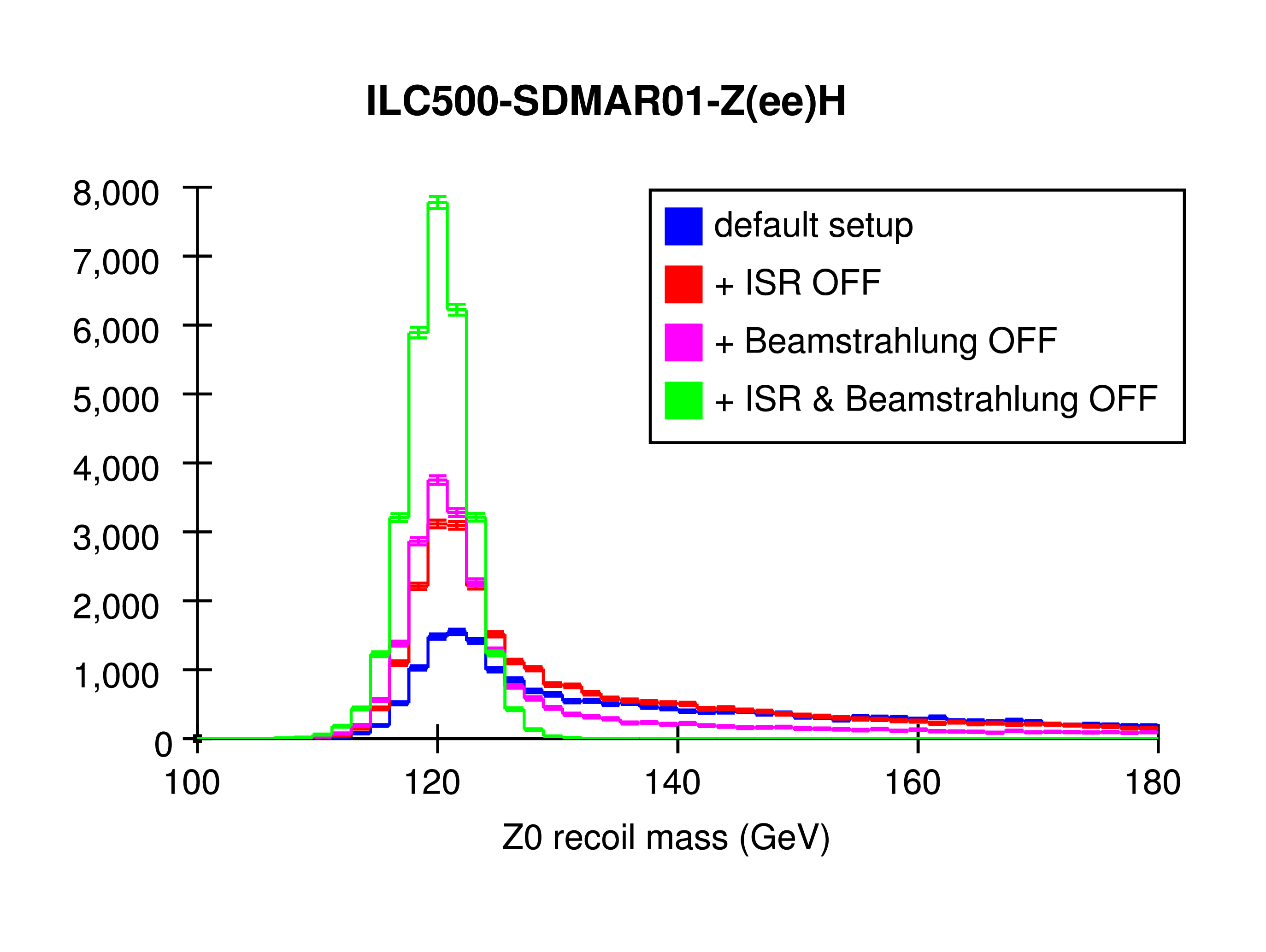}}}
{\scalebox{0.35}{\includegraphics{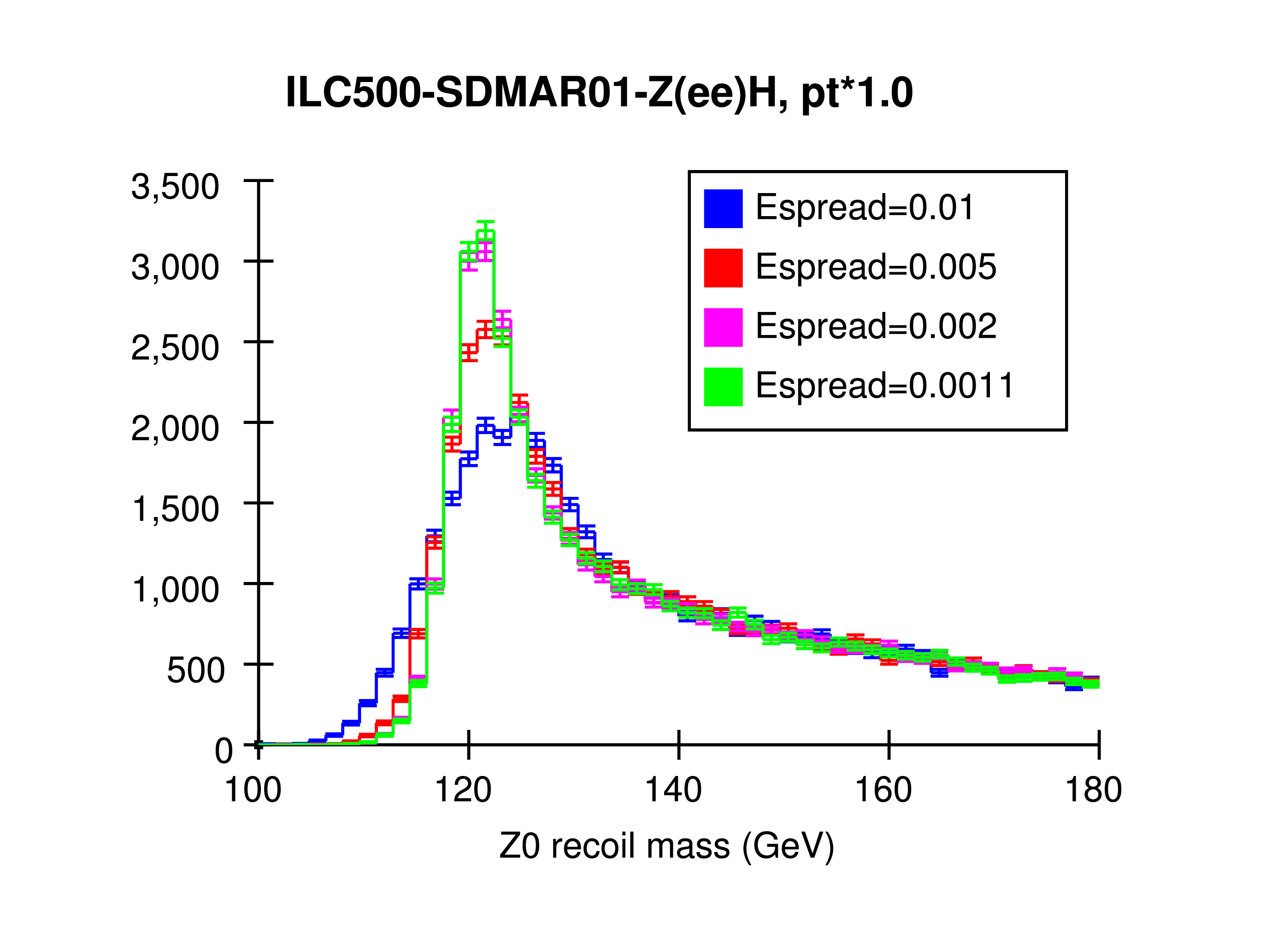}}}
\caption{\label{fig:beam} $Z^0$ recoil mass distribution for various beam setup (left), and beam energy spread (right).}
\end{figure}

As is well known, the ISR and beamstrahlung broaden the $Z^0$ recoil mass distribution and 
make an extremely long tail. The final state radiation decreases the momentum of $Z^0$ and 
makes the apparent $Z^0$ recoil mass larger than the expected value. The effects of ISR, beamstrahlung,
beam energy spread are shown in Figure~\ref{fig:beam}, assuming a 120 GeV Higgs with SD design.
As expected, the $Z^0$ recoil mass peaks at the correct Higgs mass if we turn off the
ISR and the beamstrahlung. For default detector designs the effect of beam energy spread 
becomes important for a full width larger than about 0.2\%. For the current ILC500 beam setup, 
the effect from the beam energy spread is negligible since its full width is 0.11\%.

\section{Slepton and Neutralino Mass Measurement}

The study of supersymmetry (SUSY) particles is one of the most important goals
for the future colliders. If SUSY exists in nature, there may be
light superpartners with masses of a few hundred GeV, hence, accessible
by a current proposed ILC which is expected to run at CME of 500 GeV
to 1 TeV. 

One of the ``golden'' channels to discover the SUSY particles is pair production of
slepton with each slepton decaying into lepton and lightest SUSY particle (LSP) neutralino.
The signature of these events is two energetic leptons ($e$ or $\mu$) with large missing
momentum. In this paper, we use the right-hand smuon pair production for smuon and neutralino
mass measurement, assuming the ILC operated at 500 GeV CME, 80\% right
polarization electron with integrated luminosity of 50 $fb^{-1}$. 
The smuon and neutralino masses can be determined by measuring endpoints of muon energy spectrum, 

$$
M^2_{\tilde{\mu}_R^{\pm}}  =  E^2_{cm} \bullet \frac{E_{min}\times E_{max}}{(E_{min} + E_{max})^2},~~
~M^2_{\tilde{\chi}^0_1} =  M^2_{\tilde{\mu}_R^{\pm}} \bullet  \{1 - 2 \frac{E_{min} + E_{max}}{E_{cm}}\}
 ~~~~~~~~~~~~~~~~~(1)
$$
where, $E_{cm}$ is CME, $E_{min}$ and $E_{max}$ are low and high 
muon energy end points, respectively. For Snowmass 2001 SPS\#1 point \cite{sps1}, 
the smuon and neutralino masses are 143 GeV and 96.1 GeV, respectively. 
The SUSY parameters (mSUGRA) are listed in the following,
the universal scale mass $m_0 = 100$ GeV,
the universal gaugino mass $m_{1/2} = 250$ GeV,
the trilinear coupling in Higgs sector $A_0 = -100$ GeV,
the ratio of two vacuum expectation values $tan\beta = 10$,
the Higgsino mixing parameter $sign(\mu) = +$.

\begin{figure}
{\scalebox{0.28}{\includegraphics{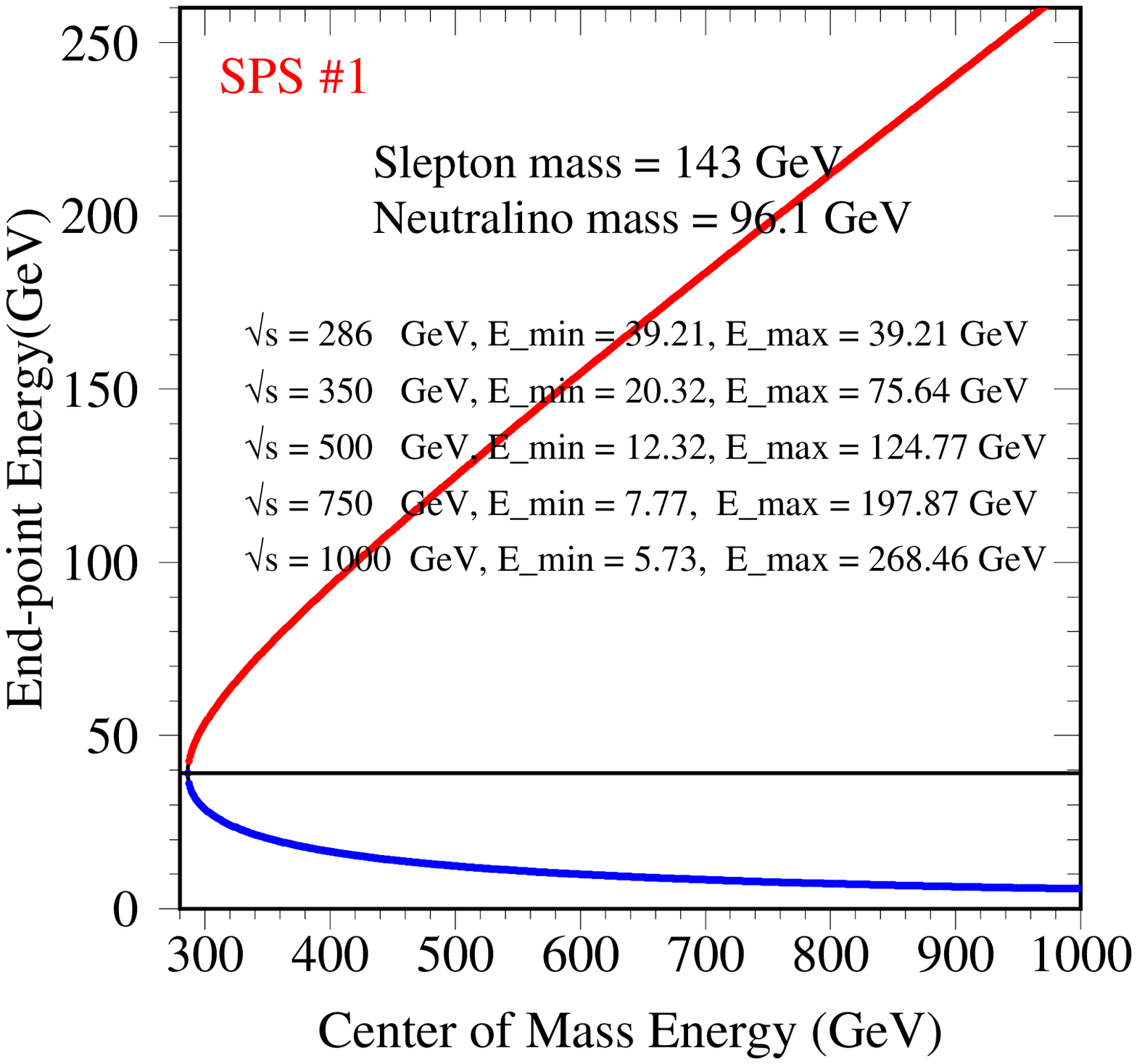}}}
{\scalebox{0.28}{\includegraphics{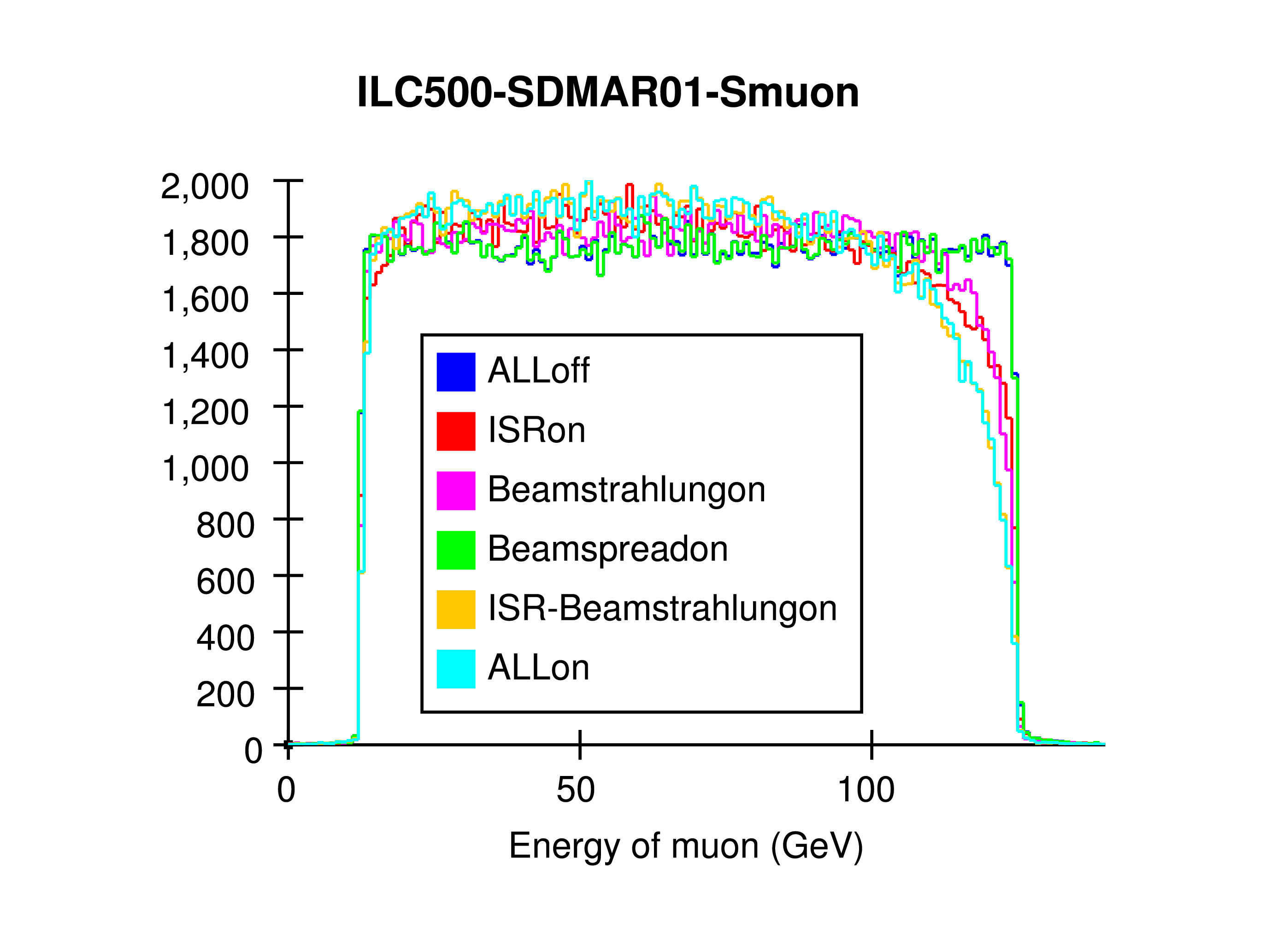}}}
{\scalebox{0.28}{\includegraphics{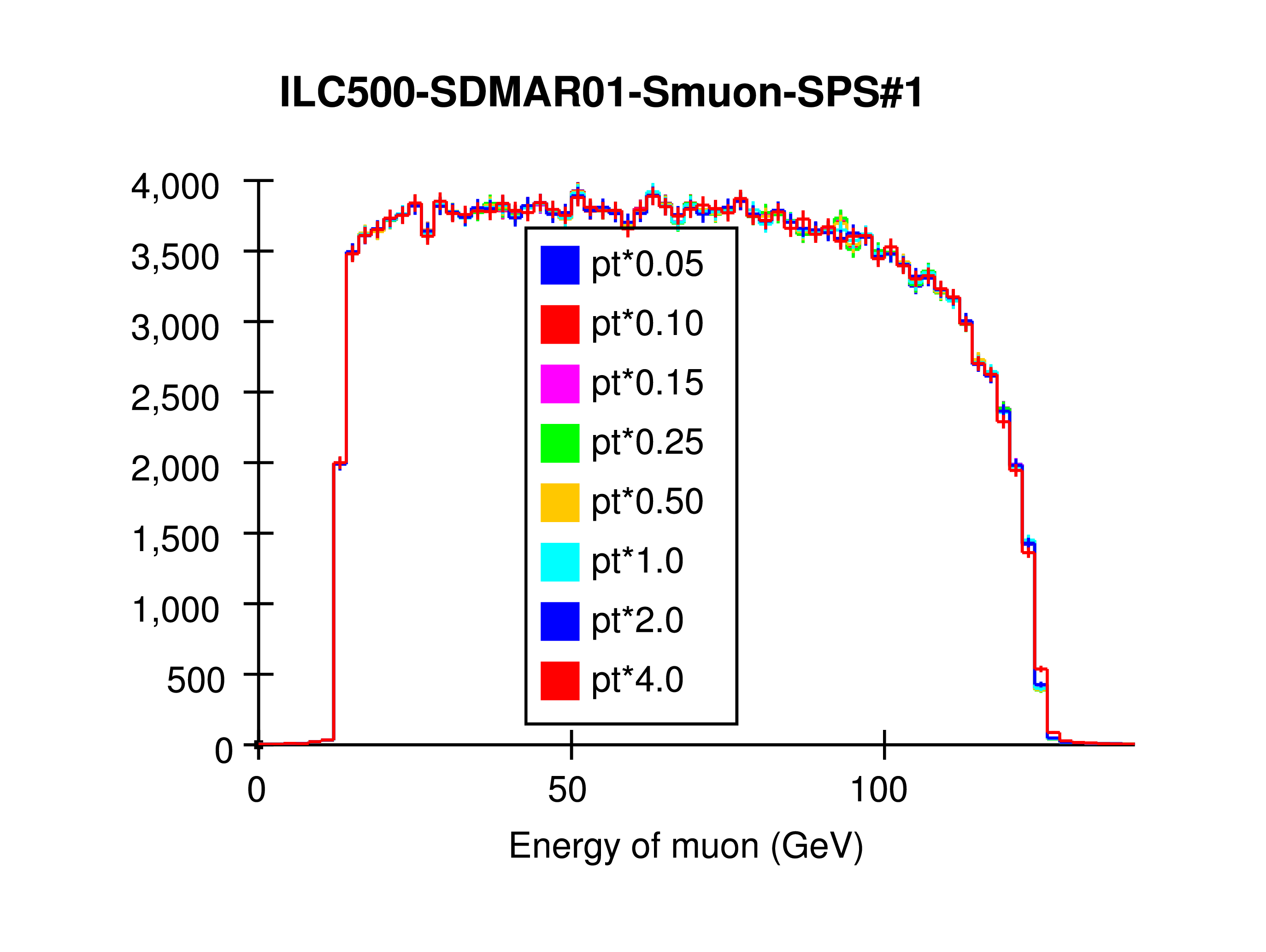}}}
\caption{\label{fig:smuon} 
Left - endpoint energy as a function of CME;
Middle - Muon energy spectra for various beam setup;
Right - Muon energy spectra for various charged track momentum resolution.
}
\end{figure}

The dependence of muon endpoint energy on the CME is
shown in the left plot of Figure~\ref{fig:smuon}. For CME
of 500 GeV, the low and high endpoint energies are 12.32 GeV and
124.77 GeV, respectively. The muon energy spectrum becomes wider with 
the increase of the CME.

The muon energy spectra for various beam setups are shown in the middle plot
of Figure~\ref{fig:smuon}. Apparently, the ISR and beamstrahlung distort the
endpoints of muon energy spectrum significantly. For instance, if we turn off
ISR, beamstrahlung and beam energy spread, the measured smuon and neutralino 
mass precisions are 260 MeV and 167 MeV, respectively, 
assuming the integrated luminosity of 50 $fb^{-1}$ and 20\% uniform random 
background contamination (20\% of signal events after selection)
over the full muon energy range. Then, we turn on the
beam energy spread with full width of 0.11\%, the measured smuon and neutralino
mass precisions are 266 MeV and 172 MeV, degraded about 3\%. If we turn on all ISR,
beamstrahlung and beam energy spread, the smuon and neutralino 
mass precisions become 420 MeV and 294 MeV, respectively. 

The dependence of the muon energy spectrum on track momentum resolution is
studied and shown in the right plot of Figure~\ref{fig:smuon}. 
Apparently, the muon energy spectra are insensitive to track momentum
resolution.

Based on equation (1), the relative error of smuon and neutralino masses 
can be written as,

$$
\frac{\Delta M_{\tilde{\mu}_R^{\pm}}}{M_{\tilde{\mu}_R^{\pm}}}
= A \bullet \sqrt{[\frac{\Delta E_{min}}{E_{min}}]^2 + [\frac{\Delta E_{max}}{E_{max}}]^2}
~~~~~~~~~~~~~~~~~~~~~(2)
$$

$$
\frac{\Delta M_{\tilde{\chi}_1^0}}{M_{\tilde{\chi}_1^0}}
= \frac{M^2_{\tilde{\mu}_R^{\pm}}}{M^2_{\tilde{\chi}_1^0}} \bullet
\sqrt{
(\frac{C}{E_{min}}-\frac{1}{E_{cm}})^2 \Delta E_{min}^2 +
(\frac{C}{E_{max}}-\frac{1}{E_{cm}})^2 \Delta E_{max}^2
}
~~~~~~~~~~~~(3)
$$
where,
$
A = \frac{E_{max} - E_{min}}{2(E_{max} + E_{min})}, ~~B = \frac{E_{max} + E_{min}}{E_{cm}},
~~C = A(1-2B)
$
. The smuon and neutralino mass errors mainly come from relative errors of $E_{min}$ and $E_{max}$
determination. Three SUSY mass pairs with high (47 GeV), medium (28 GeV) and low (6 GeV) mass differences
are considered. The selection cuts are, 
1) two muons in the final state, 
2) visible energy in the forward region less than $0.4 \times \sqrt{s}$,
3) total transverse momentum of event greater than 15 GeV to reduce most of 
$\gamma^* \gamma^*$ backgrounds if the detector can detect scattered electrons
down to about 50 mrad and  4) $|cos(\theta_\mu)| < 0.9$. 

\begin{figure}
{\scalebox{0.28}{\includegraphics{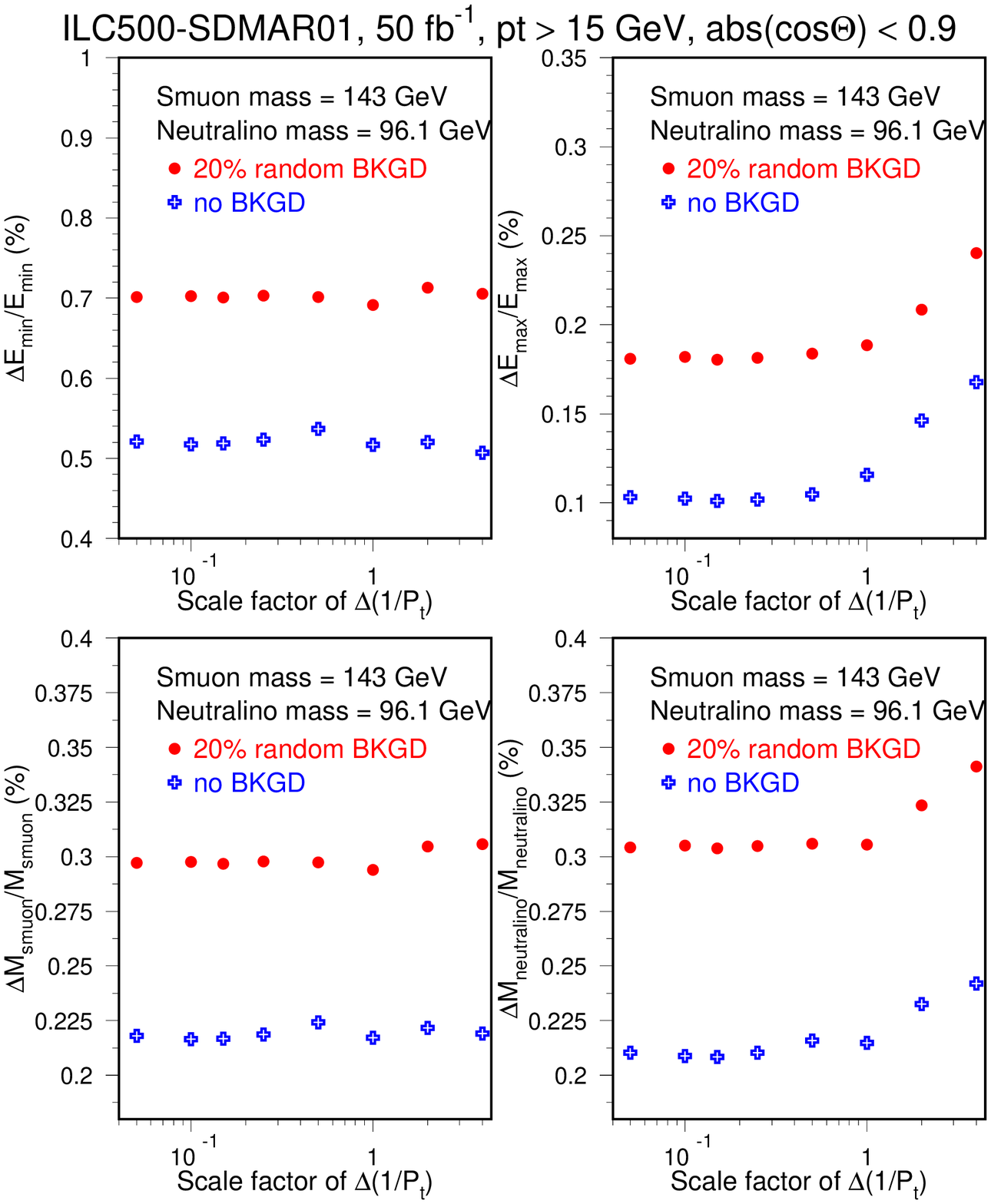}}}
{\scalebox{0.28}{\includegraphics{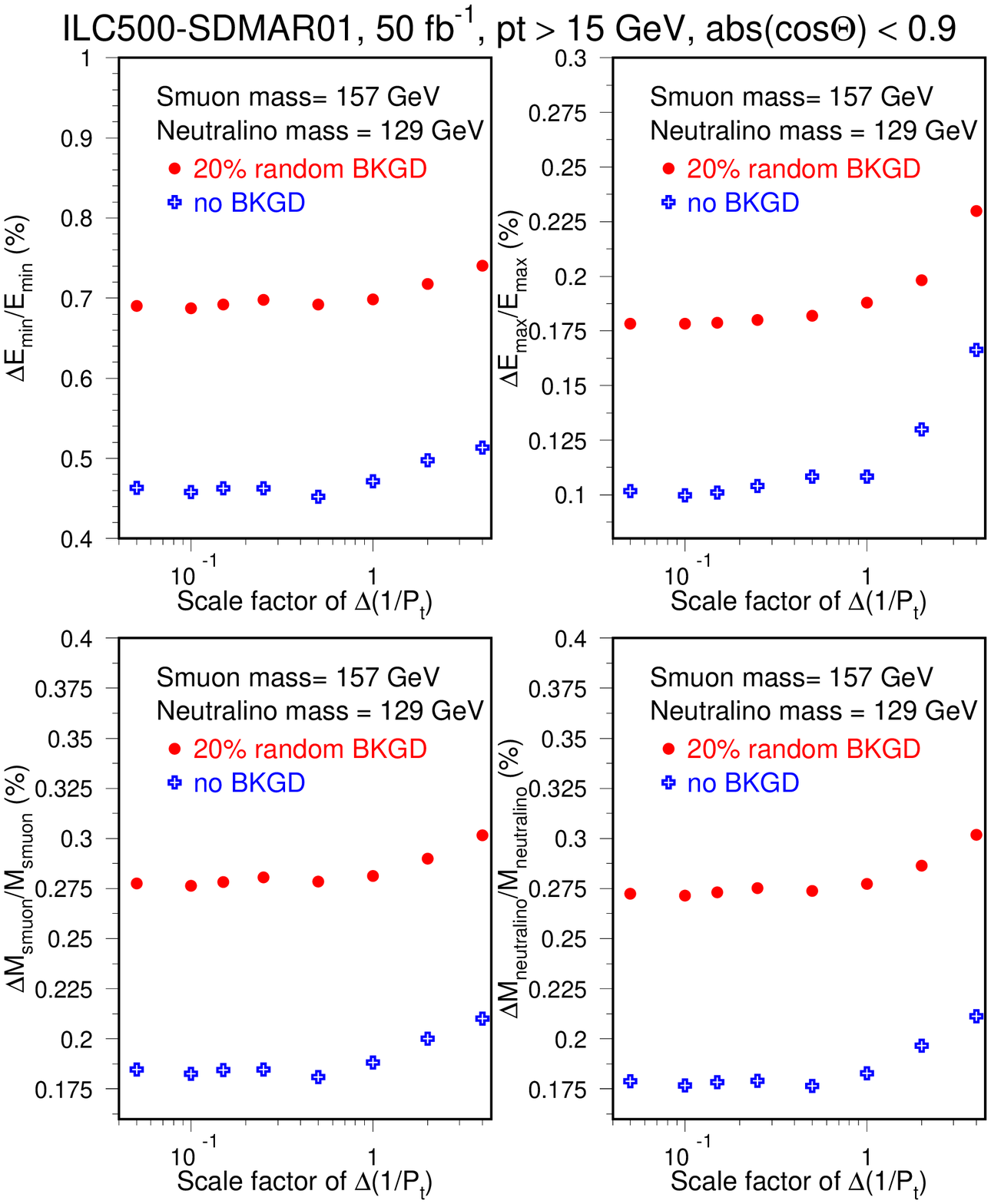}}}
{\scalebox{0.28}{\includegraphics{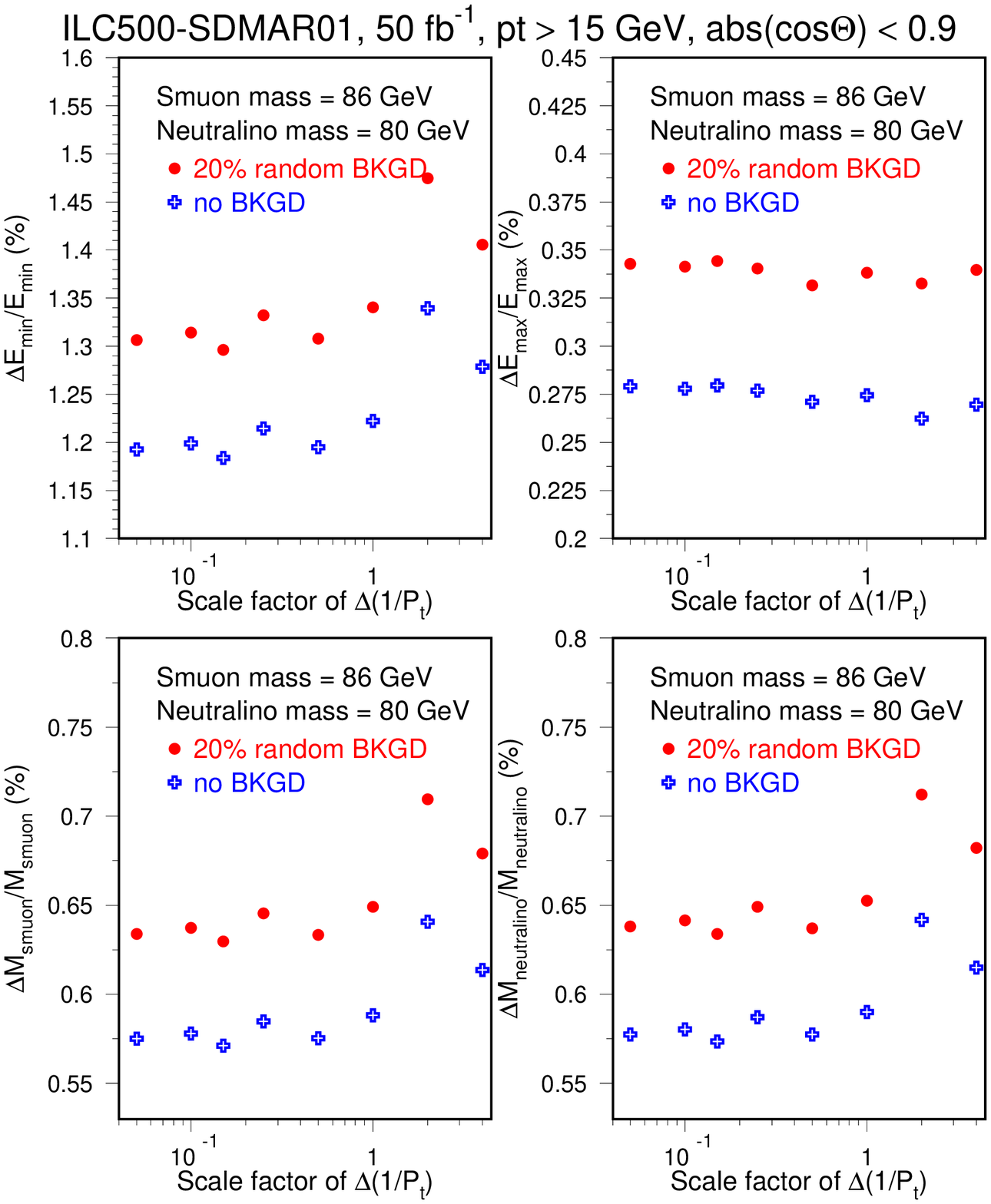}}}
\caption{\label{fig:error} 
Relative errors of $E_{min}$,  $E_{max}$,  $M_{\tilde{\mu}_R^{\pm}}$ and
$M_{\tilde{\chi}_1^0}$ versus charged track momentum resolution for
high $\Delta M = 47 ~GeV$ (left), medium $\Delta M = 28 ~GeV$ (middle),
and low $\Delta M = 6 ~GeV$ (right) mass differences.
}
\end{figure}

The relative errors of $E_{min}$,  $E_{max}$,  $M_{\tilde{\mu}_R^{\pm}}$ and
$M_{\tilde{\chi}_1^0}$ versus charged track momentum resolution for high, medium and
low mass differences are shown in Figure~\ref{fig:error}.
The smuon mass error is dominated by the relative error of the low energy endpoint $E_{min}$
although the absolute error of $E_{max}$ is larger than that of $E_{min}$.
No apparent improvement on smuon and neutralino mass precision is obtained by improving the track
momentum resolution. However, the mass precision will degrade if the
track momentum resolution get worse. The smuon and neutralino mass precisions are strongly
affected by the background contamination. The mass precisions degrade about 15-30\% 
when 20\% uniform random background (20\% of signal events after selection) is introduced.
Considering the SUSY mass error is dominated by the relative error of $E_{min}$,
it is important to reduce the background at low energy region to improve the 
mass precision.

%
%

%
%

\begin{acknowledgments}
This work is supported by the National Science Foundation 
and the Department of Energy of the United States.
\end{acknowledgments}




\end{document}